# Target Sensing Performance in Disaster-Specific ISAC Networks

Ahmet Burak Ozyurt and John S. Thompson, *Fellow, IEEE*

*Abstract*—As sixth-generation (6G) wireless technology emerges, integrated sensing and communication (ISAC) networks offer significant potential for enhancing real-time monitoring in disaster areas. However, existing ISAC approaches often fail to address the unique challenges of dynamic and cluttered disaster areas, resulting in limited sensing coverage and interruptions in sensing service. To address these limitations, this work proposes a mobile ISAC network specifically designed for disaster scenarios. By leveraging stochastic geometry, we derive closed-form expressions for sensing coverage and introduce a novel performance metric to evaluate sensing service continuity. Simulation results validate the analytical derivations and offer key insights into network design.

## I. INTRODUCTION

In recent years, the increasing frequency and severity of environmental catastrophes such as floods and earthquakes have emerged as a global concern. These disasters present critical threats to human life and necessitate innovative solutions that enable real-time monitoring and coordination to effectively support rescue operations [1]. Integrated sensing and communication (ISAC) have recently emerged as promising solutions for disaster response by simultaneously enabling real-time monitoring of survivors, hazardous leaks, and structural damages, while maintaining wireless connectivity in challenging environments [2].

Despite growing interest in ISAC, most existing solutions are limited to static or controlled scenarios, resulting in interrupted sensing service and reduced coverage in dynamic, cluttered disaster areas [3]. Although previous studies assessed the sensing coverage in controlled zones using advanced methods like the Cramer–Rao Bound (CRB) and Likelihood Ratio Test (LRT), these approaches are impractical for real-time application in disaster areas [4]–[7].

To address the aforementioned limitations, this work investigates a mobile ISAC network specifically designed for disaster scenarios. By leveraging mobile disaster response vehicles (DRVs), the proposed network can enhance both sensing coverage and sensing service continuity, two critical factors for timely and effective rescue operations [8]. Section II adopts a stochastic geometry framework to model the spatial randomness of disaster areas, with base stations (BSs) and DRVs represented as independent point processes [2], [4]. Based on this model, closed-form analytical expressions for sensing coverage are derived and a novel metric, dynamic ranging rate (DRR), is introduced in Section III to quantify sensing service continuity. Simulation results presented in Section IV validate the analytical findings and demonstrate the effectiveness of the proposed approach.

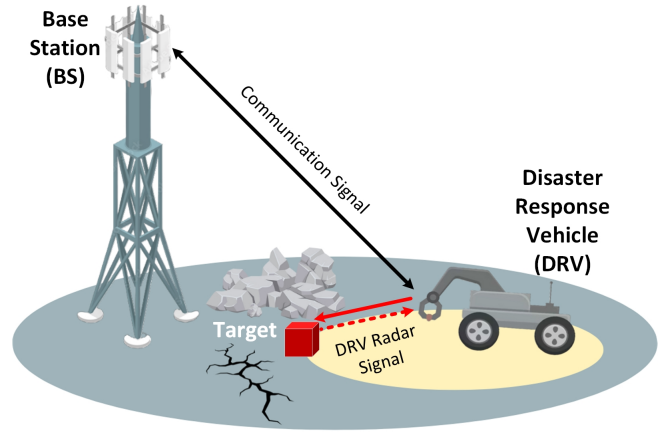

Figure 1: Network model of a disaster response vehicle (DRV) in a disaster scenario, illustrating its sensing area.

## II. SYSTEM MODEL

### A. Network Model

A disaster zone is characterized by its inherently dynamic and unpredictable nature, containing a large number of statistically random elements. As a result, employing a deterministic model for the positioning of ISAC network elements in such environments would be unrealistic. Instead, spatial point processes offer a more accurate and analytically tractable approach for disaster-specific network modeling [2]. Within this model, the positions of BSs and DRVs are represented as two independent Poisson point processes (PPPs), denoted as $\Phi_b$ and $\Phi_v$, with intensities $\lambda_b$ and $\lambda_v$, respectively. This stochastic representation enables a more flexible and realistic deployment strategy, enhancing the adaptability and efficiency of ISAC networks in disaster scenarios.

According to the preconditions outlined in the 3rd Generation Partnership Project (3GPP) Technical Report (TR) 22.870, BSs and sensor nodes support sensing capabilities in disaster areas [9]. In this disaster-specific ISAC network, BSs are characterized by low spatial density, high transmission power and static deployment, enabling them to provide extensive sensing coverage. In contrast, DRVs function as mobile units with high spatial density, lower transmission power and smaller sensing coverage, allowing for more localized operations. This hierarchical architecture establishes BSs as an umbrella layer, responsible for broad sensing coverage, while DRVs operate within the BS sensing area, enhancing sensing performance. The proposed

ISAC framework enables BSs and DRVs to simultaneously perform radar and communication functions through spatial orthogonality, thereby facilitating real-time monitoring and information dissemination in disaster scenarios, as illustrated in Figure 1.

### B. Channel Model

In network planning, received power serves as a fundamental metric for estimating sensing areas due to its simplicity, ease of measurement and effectiveness in ensuring that signals reach their intended destinations [10]. Additionally, the ISAC channel is assumed to be subject to Rayleigh fading and path loss, which are key factors influencing signal propagation in dynamic environments. Under these assumptions, the received sensing power (RSP) at the radar is given by [7]

$$S_i = \frac{P_i G_{i,t} G_{i,r} \lambda_w^2 (\sigma_t - \sigma_c)}{(4\pi)^3 d_i^{2\alpha_i}}, \quad (1)$$

where $i \in \{b, v\}$ represents the parameters associated with the BS and DRV, respectively, at a distance $d_i$; $P_i$ is the transmitting power of the sensing signal; $G_{i,t}$ and $G_{i,r}$ are the transmitting antenna gain and receiving antenna gain; $\lambda_w$ is the wavelength of the sensing signal; $\alpha_i$ is the path loss exponent, and $\sigma_t$ is the radar cross-section (RCS) of the target.

In monostatic ISAC networks, where the transmitter and receiver share the same location, as depicted in Figure 1, the transmitting and receiving antenna gains are generally equal. Consequently, it is assumed that $G_t = G_r = G$. Furthermore, since the system operates in relatively slow-moving scenarios, the target's RCS is modeled as a random variable following the Swerling type-1 model [11]. The corresponding probability density function (PDF) of the target's RCS $\sigma_t$ is given by [11]

$$f(\sigma_t) = \frac{1}{\bar{\sigma}_t} e^{-\frac{\sigma_t}{\bar{\sigma}_t}}, \sigma_t \geq 0 \quad (2)$$

where $\bar{\sigma}_t$ is the average RCS of the target. This expression of $\sigma_t$ follows a Chi-square distribution with a degree of freedom 2 [7]. Additionally, disaster zones contain obstacles and debris that can obstruct sensing signals, collectively referred to as clutter. To practically represent the disaster scenario, point clutter components are modeled as being spatially distributed around the target, centered at its location, and characterized by an RCS value of $\sigma_c$ [1].

### C. Analytical Model for DRV Sensing Area

In the context of sensing a target within a multi-layer ISAC network, it is assumed that the target is initially located within the broader sensing area of the BS. However, positioning the target within the DRV sensing area can enhance sensing service continuity. Without loss of generality, it is assumed that a typical BS is located at the origin, while a DRV is located at $\mathbf{x_s}(d_v, 0)$. Based on the RSP model presented in Section II-B, the sensing coverage of the DRV can be determined as

$$C_v = \{(x, y) \in \mathbb{R}^2 \mid S_b(d_b) = S_v(d_v)\}. \quad (3)$$

Thus, the set of equal RSP points defined in (3) forms the sensing coverage of the DRV. These points are essential for obtaining the proposed performance metric in a disaster-specific ISAC network, as discussed in Section III. For a target positioned at $(x, y) \in \mathbb{R}^2$, the respective distances from the target to the BS and the DRV are given by $d_b = \sqrt{x^2 + y^2}$ and $d_v = \sqrt{(x - d_v)^2 + y^2}$. Substituting these expressions into (3) result in

$$W(x^2 + y^2)^{\hat{\alpha}} - (x^2 + y^2 - 2d_v x + d_v^2) = 0, \quad (4)$$

where $W = \sqrt[\alpha_v]{\frac{P_v G_v^2}{P_b G_b^2}}$ and $\hat{\alpha} = \alpha_b/\alpha_v$. For realistic scenarios, it is expected that $2 < \alpha_b < 4$ and $4 < \alpha_v < 6$, which implies that $0 < \hat{\alpha} \leq 1$ [12]. Based on this relationship, the analysis is conducted for two cases depending on the path-loss exponents.

*Case 1 ($\hat{\alpha} = 1$):* In this case, a linear approximation approach is proposed for $f = (x^2 + y^2)^{\hat{\alpha}}$. To preserve the parabolic properties of the original function, it is defined as $\hat{f} = \beta(x^2 + y^2)$, where $\beta$ serves as a scaling parameter. By substituting the polar coordinate transformations $x = r\cos\theta$ and $y = r\sin\theta$ into the approximated function, the expression simplifies to $\hat{f} = \beta r^2$. To assess the accuracy of this approximation, the mean square error (MSE) incurred in approximating $f$ is considered [13]. The MSE is then expressed as

$$E(r) = |f - \hat{f}| = \sqrt{(r^{2\hat{\alpha}} - \beta r^2)^2}. \quad (5)$$

To minimize the approximation error $E(r)$, the optimal value of $\beta$ must be determined by employing the minimum mean square error (MMSE) estimator

$$\beta = \arg\min \int_0^{d_v} E(r) dr = d_v^{2(\hat{\alpha}-1)}. \quad (6)$$

The function defined in (4) can then be approximated by a closed-form circular equation, with its center at $\mathbf{x_c} = (x_r, y_r)$ and a radius $R_c$, representing the sensing area of the DRV. The center and radius are expressed as

$$\mathbf{x_c} = \left(\frac{d_v}{1 - \beta W}, 0\right), \quad R_c = \frac{\sqrt{\beta W} d_v}{1 - \beta W}. \quad (7)$$

*Case 2 ($\hat{\alpha} < 1$):* In this case, $f = (x^2 + y^2)^{\hat{\alpha}}$ is a polynomial of order less than 2, implying that the order of equation (4) is at most 2. To approximate $f$ at the point $(i, j)$, the second-order Taylor series expansion method is proposed, as given in (8a) and (8b) [10]. By substituting (8b) into (4) and performing algebraic manipulations, a generalized ellipse equation is derived to represent the sensing area of the DRV. The center of the ellipse is denoted as $\mathbf{x_e}(x_e, y_e)$ and the two semi-axes are given by $s_1$ and $s_2$. The derived equations are expressed as

$$ax^2 + 2bxy + cy^2 + 2dx + 2fy + g = 0 \quad (9)$$

$$\mathbf{x_e} = \left(\frac{cd - bf}{b^2 - ac}, \frac{af - bd}{b^2 - ac}\right) \quad (10)$$

$$s_1 = \sqrt{\frac{h}{\lambda_{\max}}}, \quad s_2 = \sqrt{\frac{h}{\lambda_{\min}}} \quad (11)$$

$$f(x,y) \approx f(i,j) + \frac{\partial f}{\partial x}\Big|_{(i,j)}(x-i) + \frac{\partial f}{\partial y}\Big|_{(i,j)}(y-j) + \frac{1}{2}\frac{\partial^2 f}{\partial x^2}\Big|_{(i,j)}(x-i)^2 + \frac{\partial^2 f}{\partial x \partial y}\Big|_{(i,j)}(x-i)(y-j) + \frac{1}{2}\frac{\partial^2 f}{\partial y^2}\Big|_{(i,j)}(y-j)^2 \tag{8a}$$

$$\approx (i^2+j^2)^{\hat{\alpha}} + 2\hat{\alpha}i(i^2+j^2)^{\hat{\alpha}-1}(x-i) + 2\hat{\alpha}j(i^2+j^2)^{\hat{\alpha}-1}(y-j) + \frac{1}{2}\Big[2\hat{\alpha}\Big((i^2+j^2)^{\hat{\alpha}-1} + (\hat{\alpha}-1)2i^2(i^2+j^2)^{\hat{\alpha}-2}\Big)(x-i)^2$$

$$+ 4\hat{\alpha}(\hat{\alpha}-1)ij(i^2+j^2)^{\hat{\alpha}-2}(x-i)(y-j) + 2\hat{\alpha}\Big((i^2+j^2)^{\hat{\alpha}-1} + (\hat{\alpha}-1)2j^2(i^2+j^2)^{\hat{\alpha}-2}\Big)(y-j)^2. \tag{8b}$$

where

$$a = W\hat{\alpha}(i^2+j^2)^{\hat{\alpha}-1} + W\hat{\alpha}(\hat{\alpha}-1)2i^2(i^2+j^2)^{\hat{\alpha}-2} - 1$$
$$b = W\hat{\alpha}(\hat{\alpha}-1)ij(i^2+j^2)^{\hat{\alpha}-2}$$
$$c = W\hat{\alpha}(i^2+j^2)^{\hat{\alpha}-1} + W\hat{\alpha}(\hat{\alpha}-1)2j^2(i^2+j^2)^{\hat{\alpha}-2} - 1$$
$$d = W\hat{\alpha}i(i^2+j^2)^{\hat{\alpha}-1} + d_v$$
$$f = W\hat{\alpha}j(i^2+j^2)^{\hat{\alpha}-1}$$
$$g = W(i^2+j^2)^{\hat{\alpha}} - d_v^2$$
$$h = g - ax_e^2 - 2bx_e y_e - cy_e^2$$
$$\lambda = [(a+c) \pm \sqrt{(a-c)^2 + 4b^2}]/2. \tag{12}$$

## III. SENSING SERVICE CONTINUITY

With the emergence of ISAC technology, integrating sensing with multiple network elements has become essential for achieving accurate and large-scale sensing in various scenarios [8]. However, multiple network elements complicate sensing service continuity, causing higher signalling overhead. To address this, we propose a novel performance metric, the stochastic geometry-based dynamic ranging rate (DRR), specifically designed to evaluate sensing service continuity in disaster-specific ISAC networks. This metric characterizes the process whereby a target, initially detected by a BS, undergoes further localization by mobile sensing nodes to enhance accuracy. Consequently, the DRR, denoted by $\xi$, is defined as

$$\xi = \xi_r \times \mathbb{P}(\kappa > \tau), \tag{13}$$

where the dynamic ranging repetition rate $\xi_r$ denotes the number of times a target initially located within the BS sensing area enters the DRV sensing area per unit time, due to DRV mobility. The term $\mathbb{P}(\kappa > \tau)$ represents the probability that the target's dwell time $\kappa$ within the DRV sensing area exceeds the pulse repetition interval (PRI) $\tau$ [14]. This probability indicates the likelihood of sufficient signal integration to enable effective target sensing.

Movement patterns in disaster zones exhibit complex spatio-temporal correlations, complicating accurate modeling [2]. Thus, we adopt an improved random waypoint (RWP) mobility model, which effectively captures node mobility with analytical simplicity [10]. The dynamic ranging repetition rate $\xi_r$ is defined as the ratio of the expected number of ranging events $E[N]$, within one movement period, to the expected duration of that period $E[T]$, expressed as $\xi_r = E[N]/E[T]$.

In the RWP mobility model, the trajectory during the $k$-th movement period is defined by two successive waypoints, $\mathbf{X}_{k-1}$ and $\mathbf{X}_k$. The number of dynamic ranging events is obtained by counting intersections between the target's relative trajectory and the DRV sensing area, denoted as $L(\mathbf{X}_{k-1}, \mathbf{X}_k)$. Consequently, the expected dynamic ranging repetition rate for a target is derived as

$$\mathbb{P}_r = \int_0^\infty \int_0^\infty \frac{2|L(\mathbf{X}_{k-1}, \mathbf{X}_k)|R_c}{|\mathcal{A}|} f_{|L(\mathbf{X}_{k-1}, \mathbf{X}_k)|}(l) f_{R_c}(r) dl dr$$
$$= \frac{2}{|\mathcal{A}|} E[|L(\mathbf{X}_{k-1}, \mathbf{X}_k)|] E[R_c]$$
$$= \frac{\sqrt{W}}{1-W} \cdot \frac{1}{2|\mathcal{A}|\sqrt{\lambda_b \lambda_v}}, \tag{14}$$

where $f_{|L(\mathbf{X}_{k-1}, \mathbf{X}_k)|}(l)$ denotes the PDF of the DRV's transition length modeled as a Rayleigh distribution. Thus, the expected transition length is $E[|L(\mathbf{X}_{k-1}, \mathbf{X}_k)|] = 1/(2\sqrt{\lambda_v})$. Furthermore, $f_{R_c}(r)$ represents the PDF of the DRV's sensing radius. Hence, the expected number of dynamic ranging events for a DRV is given by $E[N] = \lambda_v |\mathcal{A}| \mathbb{P}_r$, where $|\mathcal{A}|$ denotes the area of interest. Moreover, the expected movement period is given by $E[T] = E[T_s] + E[T_t]$, where $E[T_s]$ and $E[T_t]$ denote the mean pause time and the mean transition time, respectively. Specifically, the mean transition time is expressed as $E[T_t] = E[|L(\mathbf{X}_{k-1}, \mathbf{X}_k)|/u]$, with $u$ representing the DRV's velocity. Therefore, the dynamic ranging repetition rate is expressed as

$$\xi_r = \frac{\sqrt{W}}{1-W} \frac{\lambda_v u}{\sqrt{\lambda_b} + 2\sqrt{\lambda_b \lambda_v} u E[T_s]}. \tag{15}$$

According to (13), a dynamic ranging event occurs when the target's dwell time $\kappa$ within the DRV's sensing area exceeds the PRI $\tau$. Additionally, the expected trajectory length of the target inside a DRV sensing area with radius $R_c$ is given by $L_c(R_c) = \frac{\pi}{2} R_c$. Substituting $R_c$ from (7), directly relates to the DRV sensing area's characteristics.

Furthermore, by leveraging the stochastic geometry properties of distance distributions in a PPP, the probability that the target's dwell time $\kappa$ within the DRV sensing area exceeds the PRI $\tau$ is given by

$$P(\kappa \geq \tau) = \exp\left\{ -\frac{4\lambda_b}{\pi} \frac{(1-W)^2}{W} u^2 \tau^2 \right\}. \tag{16}$$

By substituting (15) and (16) into (13), we obtain the closed-form expression for the dynamic ranging rate as follows

$$\xi = \frac{\sqrt{W}}{1-W} \frac{\lambda_v u}{\sqrt{\lambda_b} + 2\sqrt{\lambda_b \lambda_v} u E[T_s]} \times e^{-\frac{4\lambda_b}{\pi} \frac{(1-W)^2}{W} u^2 \tau^2}. \tag{17}$$

## IV. NUMERICAL RESULTS

This section presents analytical results for the DRV sensing area and DRR performance metric derived previously, validated by Monte-Carlo simulations. Unless otherwise specified, we consider BSs and DRVs distributed as two independent homogeneous PPPs with intensities $\lambda_v = 2\lambda_b = 1$ node/km$^2$. The default system parameters are set as $P_b = 46$ dBm, $P_v = 30$ dBm, $G_b = 14$ dBi, $G_v = 5$ dBi, $E[T_s] = 0.5$ s and $u = 1.4$ m/s.

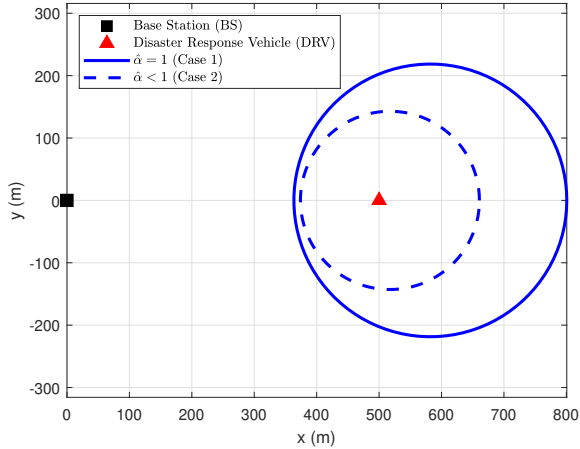

Figure 2: Disaster response vehicle (DRV) sensing coverage model with a base station (BS) positioned at the origin.

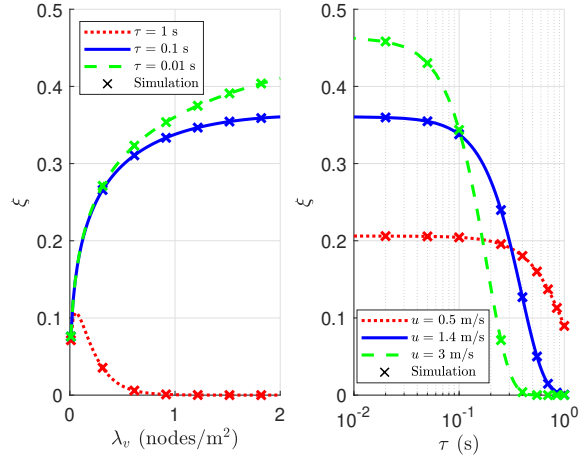

Figure 3: (a) DRR vs. DRV intensity for different PRI values (left); (b) DRR vs. PRI for different DRV velocities (right).

Figure 2 illustrates the sensing coverage of the DRV for different path-loss exponents, with a BS positioned at the origin and the DRV located at (500 m, 0). To assess the approximation methods discussed in Section II-C, we compare Case 1 (linear approximation) and Case 2 (ellipse approximation). It is observed that the sensing coverage in Case 1 is notably larger than in Case 2, even though both scenarios share the same BS coverage. This difference arises primarily due to the distinct path-loss exponents, namely a smaller exponent in Case 1 enhances the sensing coverage compared to Case 2. While both methods use identical transmit power, the higher path-loss exponent in Case 2 significantly restricts signal propagation, leading to a reduced sensing area.

Figure 3 illustrates the DRR $\xi$ as a function of DRV intensity $\lambda_v$ and PRI $\tau$. The results indicate that DRR increases with higher DRV density and velocity. Specifically, as $\lambda_v$ increases, the probability of targets falling within DRV sensing coverage also rises, enhancing dynamic ranging events. Conversely, DRR declines as $\tau$ increases due to the inherent trade-off between detection range and velocity resolution. Larger PRI values result in reduced resolution, compromising detection and tracking performance. Although shorter PRI values yield improved Doppler accuracy, they limit the maximum detection range. Notably, beyond $\tau$= 0.1 s, performance becomes inconsistent and unreliable, suggesting that excessively large PRI values degrade system effectiveness. Thus, optimal selection of PRI is crucial for maintaining robust ranging accuracy. Furthermore, analytical results closely match simulation outcomes, validating the proposed mathematical model and offering valuable insights for optimization in ISAC-based disaster scenarios.

## V. CONCLUSION

This work proposes a stochastic geometry framework for disaster-specific ISAC networks, deriving closed-form expressions for sensing coverage and introducing the DRR metric for evaluating sensing service continuity. Monte Carlo simulation results validate the analytical findings, demonstrating that increased network density and DRV velocity enhance dynamic ranging performance. However, excessively large PRI values degrade performance, revealing a critical threshold for effective system optimization.